\documentclass[aps,superscriptadress,preprint]{revtex4}
\usepackage{graphicx}
\usepackage{amsmath}

\begin{document}
\title{Quark deconfinement phase transition for improved quark mass
density-dependent model}

\author{Chen Wu$^1$
and Ru-Keng Su$^{1,2,3}$\footnote{rksu@fudan.ac.cn}} \affiliation{
\small 1. Department of Physics, Fudan University,Shanghai 200433, P.R. China\\
\small 2. CCAST(World Laboratory), P.O.Box 8730, Beijing 100080,
P.R. China\\
\small 3. Center of Theoretical Nuclear Physics,\\
National Laboratory of Heavy Ion Collisions, Lanzhou 730000,
P.R.China}
\begin{abstract}
By using the finite temperature quantum field theory, we calculate
the finite temperature effective potential  and extend the improved
quark mass density-dependent model to finite temperature. It is
shown that this model can not only describe the saturation
properties of nuclear matter, but also explain the quark
deconfinement phase transition successfully. The critical
temperature is given and the effect of $\omega$- meson is addressed.
\end{abstract}

\maketitle

\section{Introduction}
Quark- meson coupling(QMC) model suggested by Guichon [1] is a
famous hybrid quark meson model, which can describe the saturation
properties of nuclear matter and many other properties of nuclei
successfully. In this model, the nuclear system was suggested as a
collection of MIT bag, $\omega$- meson and $\sigma$- meson and the
interactions between quarks and mesons are limited within the MIT
bag regions because the quark cannot escape from the MIT bag. This
model had been extended by many authors, for example, including
$\rho$ meson to discuss the neutron matter [2], adding hyperons to
study the strange hadronic matter [3], suggesting the density-
dependent vacuum energy to investigate the effect of environment [4]
and etc. [5].

Although the QMC model is successful in describing many physical
properties of  nuclear systems, but as was pointed in our previous
paper [6], it has two major shortcomings: (1) It is  a permanent
quark confinement model because the MIT bag boundary condition
cannot be destroyed by temperature and density. It cannot  describe
the quark deconfinement phase transition. (2) It is difficult to do
nuclear many-body calculation beyond mean field approximation(MFA)
by means of QMC model, because we cannot find the free propagators
of quarks and mesons  easily. The reason is that the interactions
between quarks and mesons are limited within the bag regions, the
multireflection of quarks and mesons by MIT bag boundary must be
taken into account for getting the free propagators. These two
shortcomings actually inherited from the MIT bag.

To overcome these two shortcomings, in our previous paper [6], we
suggested an improved quark mass density- dependent(QMDD) model. The
QMDD model was suggested by Fowler, Raha and Weiner [7] many years
ago. According to the QMDD model, the masses of u, d, s quarks and
corresponding antiquarks are given by
\begin{eqnarray}
m_{q} = \frac{B}{3n_{B}}(i = u,d,\bar u,\bar d)   \label{mq1}
\end{eqnarray}
\begin{eqnarray}
m_{s,\bar s } = m_{s0}+\frac{B}{3n_{B}}     \label{mq2}
\end{eqnarray}
where $m_{s0}$ is the current mass of the strange quark, $B$ is the
bag constant, $n_{B}$ is the baryon number density
\begin{eqnarray}
n_B=\frac{1}{3}(n_u+n_d+n_s),
\end{eqnarray}
 $n_u,n_d,n_s$ represent the density of u quark, d quark, and s
quark, respectively. As was explained and proved in Ref. [8, 9], the
basic hypothesis (1), (2) correspond to a confinement mechanism of
quarks because  when volume $V\rightarrow \infty $, $n_B \rightarrow
0 $, and then $m_q \rightarrow \infty $, quark will be confined.
Using QMDD model, many authors investigated the dynamical and
thermodynamical properties of strange quark matter and strange star
and found that the results given by QMDD model are nearly the same
as those obtained  in MIT bag model [8-16]. But a nice advantage for
QMDD model is that the MIT bag boundary condition has been given up.
The ansatz Eq. (1) is employed to replace the MIT bag boundary.

But QMDD model still has two shortcomings: (1) It is still an ideal
quark gas model. No interactions  between quarks  exist except a
confinement ansatz Eqs. (1-2). (2) It still cannot explain the quark
deconfinement phase transition and give us a correct phase diagram
as that given by lattice QCD because when $n_B \rightarrow 0$,$m_q
\rightarrow \infty $,  $T\rightarrow \infty $ [8]. To overcome the
difficulty(2), we introduced a new ansatz that $m_q$ is not only a
function of density, but also depends on temperature T [8,12,15], we
suggested
\begin{eqnarray}
B=B_0[1-(T/T_C)^2], 0\leq T \leq T_C, \\
B=0\ \ \ \ \ \ \ \ \ \ \ \ \ \ ,  T > T_C
\end{eqnarray}
and extended the QMDD model to a quark mass density- and
temperature- dependent model(QMDTD) model. The ansatz(4) guarantees
that $m_q \rightarrow 0$ when $T \rightarrow T_C$. It changes the
permanent confinement mechanism(MIT bag) to  a nonpermanent
confinement mechanism (Friedberg-Lee(FL) soliton bag) in the QMDTD
model. Since the vacuum density B equals to the different value
between the local false vacuum minimum and the absolute real vacuum
minimum of nonlinear scalar field in FL model, we introduced a
nonlinear scalar field to improve the QMDD model in Refs. [17, 22]
and changed the ad-hoc ansatz(4) to an output B(T) curve which can
be calculated from the improved QMDD model.

To improve the shortcoming(1) of QMDD model, we introduced the
$\omega-$ meson and $\sigma-$ meson in QMDD model to mimic the
repulsive and attractive interactions between quarks in Refs. [6,
23]. The interaction between quarks and nonlinear $\sigma$ field
forms a FL soliton bag [17, 22].  The $qq\omega$  and $qq\sigma$
interaction guarantees that we can get valid  saturation properties,
the equation of state and the compressibility  of nuclear matter
[6]. But the deconfinement phase transition has not yet been studied
in Ref. [6]. This motivates us to study the deconfinement properties
of the improved QMDD(IQMDD) model in this paper. We would like to
emphasize that this is the basic important advantage for IQMDD
model, because the saturation properties can be explained not only
 by IQMDD model but also by QMC model. The reason for the
 explaination of quark
 deconfinement  by IQMDD model is that
MIT boundary constraint has been dropped   and interactions between
quark and mesons have been extended to the whole space.  Since
instead of the MIT bag in QMC model, a FL soliton bag is introduced
in IQMDD model, we can use our model to discuss the quark
deconfinement phase transition. The spontaneous breaking symmetry of
nonlinear $\sigma$ field will be restored and the soliton bag will
disappear at critical temperature [24-27].

The organization of this paper is as follows. In the next section,
we give the main formulae of the IQMDD model and effective potential
 at finite temperature. The soliton solutions of
IQMDD model at different temperature and other numerical results
are presented in the third section. The last section contains a
summary and discussions.

\section{The IQMDD model at zero and finite temperature}
The effective Lagrangian density of the IQMDD model is given by
\begin{eqnarray}
\mathcal{L}=\overline\psi[i\gamma^{\mu}\partial_{\mu}-
m_{q}-f\sigma-g\gamma^{\mu}\omega_{\mu}]\psi \hskip 0.8in \nonumber\\
+\frac{1}{2}\partial_{\mu}\sigma \partial^{\mu}\sigma -U(\sigma)
-\frac{1}{4}F_{\mu\nu}F^{\mu\nu}+ \frac{1}{2}m_{\omega}^2
\omega^{\mu}\omega_{\mu},
\end{eqnarray}
where
$F_{\mu\nu}=\partial_{\mu}\omega_{\nu}-\partial_{\nu}\omega_{\mu}$,
$\psi$ represents the quark field, $m_{q}=\frac{B}{3n_{B}}$ is the
mass of $u(d)$ quark, the $\sigma$  and $\omega$ field are not
dependent on time, $f$ is the coupling constant between the quark
field $\psi$ and the scalar meson field $\sigma$, $g$ is the
coupling constant between the quark field $\psi$ and the vector
meson field $\omega_{\mu}$, $U(\sigma)$ is the self interaction
potential for $\sigma$ field. We omit the contribution of the s
quark  and consider the nuclear system only in this paper. The
potential field $U(\sigma)$ is chosen as [18]
\begin{eqnarray}
U(\sigma)=\frac{a}{2!}\sigma^2+\frac{b}{3!}\sigma^3+\frac{c}{4!}\sigma^4+B,
\end{eqnarray}
\begin{equation} \label{e7}
b^2 > 3ac
\end{equation}
The condition (\ref{e7}) ensures that the absolute minimum of
$U(\sigma)$ is at $\sigma = \sigma_{v} \ne 0$. The potential
$U(\sigma)$ has two minima: one is the absolute minimum $\sigma_v$
\begin{eqnarray}
\sigma_v=\frac{3|b|}{2c}\left[1+\left[1-\frac{8ac}{3b^2}\right]^{\frac
1 2}\right],
\end{eqnarray}
it  corresponds to the physical  vacuum,  and the other is at
$\sigma_0=0$, it represents a metastable local false vacuum. We take
$U(\sigma_v)=0$ and the bag constant $B$ can be expressed as
\begin{eqnarray}
-B=\frac{a}{2!}\sigma^2_v+\frac{b}{3!}\sigma^3_v+\frac{c}{4!}\sigma^4_v.
\end{eqnarray}

From Eq. (6), we obtain the equation of motion for quark as
\begin{equation}
(i \gamma^{\mu} \partial_{\mu} -m_{q}- f \sigma - g
\gamma^{\mu}\emph{} \omega_{\mu})\psi=0,
\end{equation}

and the equations for the scalar meson field and vector meson field
as
\begin{eqnarray}\label{euler-eq1}
\partial_{\mu}\partial^{\mu}\sigma + \frac{dU(\sigma)}{d\sigma}=
-f\bar{\psi}\psi,
\end{eqnarray}
\begin{equation}
\partial_{\nu}
F^{\nu\mu}+ m_\omega \omega^2=g\bar{\psi}\gamma^{\mu}\psi.
\end{equation}
respectively. Using an approximation as that of the QMC model, we
replace $\sigma(\mathbf{r},t)\rightarrow \sigma(r),$
$\omega_{\mu}(\mathbf{r},t) \rightarrow \delta_{\mu0}\omega(r)$ and
consider a fixed occupation number of valence quarks (3 quarks for
nucleons, and quark-antiquark pair for mesons)only.  In the
following, we will discuss the ground state solution of the system.
The Hamiltonian density is
\begin{eqnarray}
\mathcal{H}=\psi^{+}[\frac{1}{i}\vec{\alpha}\cdot\vec{\nabla}+
\beta(m_{q}+f\sigma)+g\omega]\psi+\frac{1}{2}\Pi_{\sigma}^{2}
\nonumber\\+\frac{1}{2}(\nabla\sigma)^{2}+U(\sigma)-
\frac{1}{2}(\nabla \omega)^{2}-\frac{1}{2}m_\omega^2 \omega_\mu
\omega^\mu.
\end{eqnarray}
where $\vec{\alpha}$ and $\beta$ are the  Dirac matrices£¬
$\Pi_{\sigma}$ is conjugate field of the scalar meson field. One can
construct a Fock space  of quark states and expand the operator
$\psi$ in terms of annihilation and creation operators on the space
with spinor function $\varphi^{\pm}_n$, which satisfies the Dirac
equation [17]:
\begin{equation}
[\vec{\alpha}\cdot \vec{p}+
\beta(m_{q}+f\sigma)+g\omega]\varphi^{\pm}_{n}=\pm
\epsilon_{n}\varphi^{\pm}_{n}.
\end{equation}

The functions $\varphi_{n}$ satisfies the normalized condition
$\int\varphi_{n}^{+}\varphi_{n}d^{3}r=1$. From Eq. (14), the total
energy of the system is given by
\begin{equation}
E(\sigma,\omega)=\sum_{n}\epsilon_{n}+ \int
[\frac{1}{2}(\nabla\sigma)^{2}+U(\sigma)- \frac{1}{2}(\nabla
\omega)^{2}-\frac{1}{2}m_\omega^2 \omega^2]d^{3}r.
\end{equation}

Substituting Eq. (15) into Eqs. (12, 13), and using the variational
principle under the spherical symmetric condition, we find when
$\sigma$ and $\omega$ satisfy the follow equations
$$
-\nabla_{r}^{2}\sigma+\frac{dU(\sigma)}{d\sigma}=-f\sum_{n}\bar{\varphi_{n}}\varphi_{n},$$
\begin{equation}
-\nabla_{r}^{2}\omega+m_\omega^2
\omega=g\sum_{n}\varphi_{n}^{+}\varphi_{n}.
\end{equation}
respectively where $\overline\varphi_n = \varphi_n^{+}\gamma_0$, we
have the minimum of $E(\sigma,\omega)$.

We discuss the ground state solution of the system now. The quark
spinor in the lowest state is assumed [17-19]:
\begin{equation}
\varphi=\left(%
\begin{array}{c}
  u(r) \\
  i(\frac{\vec{\sigma} \cdot \vec{r}}{r})v(r) \\
\end{array}%
\right)\chi_{m},\  \ \chi_{m}=\left(%
\begin{array}{c}
  1 \\
  0 \\
\end{array}%
\right) or
\left(%
\begin{array}{c}
  0 \\
  1 \\
\end{array}%
\right),
\end{equation}
where $\vec{\sigma}$ are the Pauli matrices. Substituting Eq. (18)
into Eq. (15), we get the equations of spinor components $u$ and $v$
as
$$
\frac{du(r)}{dr}=-[\epsilon+m_{q}+f\sigma(r)-g\omega]v(r),$$
\begin{equation}
\frac{dv(r)}{dr}=-\frac{2}{r}v(r)+[\epsilon-m_{q}-f\sigma(r)-g\omega]u(r).
\end{equation}
respectively. The normalized condition then reads as
$4\pi\int^{\infty}_{0}[u^{2}(r)+v^{2}(r)]r^{2}dr=1$.

From  Eq. (17) and  Eq. (18), after summation of the quark states,
we obtain the equation of motion of the $\sigma,\omega$ field
\begin{equation}
\frac{d^{2}\sigma}{dr^{2}}+ \frac{2}{r}\frac{d\sigma}{dr}-
\frac{dU(\sigma)}{d\sigma}=Nf(u^{2}-v^{2}),
\end{equation}
\begin{equation}
\frac{d^{2}\omega}{dr^{2}}+ \frac{2}{r}\frac{d\omega}{dr}-
m_\omega^2 \omega=-Ng(u^{2}+v^{2})\equiv F(r).
\end{equation}
where, the number of quarks is $N=3$ for baryons and $N=2$ for
mesons. In the following discussions, we only consider the case
$N=3$. To get a self-consistent solution of Eqs. (19, 20, 21),  we
select the boundary conditions for quark field and $\sigma$ field as
[17, 22]
\begin{eqnarray}
v(r=0)=0,  u(r=\infty)=0, \nonumber\\
\sigma^\prime (r=0)=0, \sigma(r=\infty)=\sigma_{v}.
\end{eqnarray}
respectively. Noting that the $r\rightarrow \infty$ asymptotic
behavior of the $\omega$ field given by Eq. (21) tends to an
exponent decay wave because $F(r\rightarrow \infty) \rightarrow 0$
for a soliton bag, we can find the corresponding Green function
$G(r,r^\prime)$ easily and obtain the $\omega$ field by integral
[20]:
\begin{eqnarray}
\omega(r)=\int_0^\infty r^{\prime 2}dr^\prime F(r^\prime)G(r,
r^\prime)
\end{eqnarray}
The numerical results of $u(r), v(r), \sigma(r)$ and $\omega(r)$
will be  shown in next section.

In order  to study the deconfinement phase transition, we turn to
extend IQMDD model to finite temperature.   The appropriate
framework is the finite temperature quantum field theory. The finite
temperature effective potential plays a central role within this
framework. Under mean field approximation, the meson field operators
can be replaced by their expectation values,
$\omega_{\mu}\rightarrow \bar\omega_{\mu}=\delta_{\mu
0}\bar\omega=\delta_{\mu 0}\frac{g}{m_\omega^2}\rho_{B0}$[20, 21].
Using the method of Dolan and Jackiw [24], up to one-loop
approximation, the effective potential reads:
\begin{eqnarray}\label{potential0}
V(\sigma;T; \mu; V_\omega)=U(\sigma)+V_B(\sigma;T)+V_F(\sigma;T;
\mu; V_\omega),
\end{eqnarray}
where
\begin{eqnarray}
V_\omega=g \bar\omega=\frac{g^2}{m_\omega^2}\rho_{B0}
\end{eqnarray}
is the contribution of $\omega$-field, $\rho_{B0}$ is saturation
density of nuclear  matter, T is temperature and

\begin{eqnarray}\label{potential1}
V_B(\sigma;T)=\frac{T^4}{2\pi^2} \int^{\infty}_0 dx x^2 \mathrm{ln}
\left( 1-e^{-\sqrt{(x^2+ m_{\sigma}^2/T^2)}} \right),
\end{eqnarray}
\begin{eqnarray}\label{potential2}
V_F(\sigma;T; \mu; V_\omega)=-12\sum_n \frac{T^4}{2\pi^2}
\int^{\infty}_0 dx x^2 \mathrm{ln} \left( 1+e^{-(\sqrt{(x^2+
m_{qn}^2/T^2)}- \mu_n/T+  V_{\omega}/T)} \right),
\end{eqnarray}
where the minus sign of Eq. (27) is the consequence of Fermi-Dirac
statistics. The degenerate factor 12 comes from: 2(particle and
antiparticle), 2(spin), 3(color). $m_{\sigma}$ and $m_q$ are the
effective masses of the scalar field $\sigma$ and the quark field
respectively:
\begin{eqnarray}
m^2_{\sigma}&=& a+b \sigma(T)+\frac{c}{2} \sigma^2(T)\\
 m_q &=&\frac{B(T)}{n_q}+f \sigma(T).
\end{eqnarray}

We see from Eqs. (24-27) that the scalarlike interaction
$f\psi^+\sigma\psi$ gives contribution to effective masses of quark
and $\sigma$\- meson and then forms a confined soliton bag, the
vectorlike interaction $g\psi^+\gamma^{\mu}\omega_{\mu}\psi$ gives
contribution to an effective chemical potential of quarks. In fact,
this finite temperature effective potential for FL model had been
calculated by many others authors including us [22, 25-27] , but
without $\omega$- field.

In the  soliton bag  model, the finite temperature vacuum energy
density $B(T)$ is defined as
\begin{eqnarray}\label{bag}
B(T; \mu; V_\omega)=V(\sigma_0;T; \mu; V_\omega)-V(\sigma_v;T; \mu;
V_\omega).
\end{eqnarray}
It is the different from the values at the perturbative false vacuum
state and the values at the physical real vacuum state of the finite
temperature effective potential. At critical temperature $ T_{C}$ of
quark deconfinement phase transition, $B$ equal to zero: $B(T_C)=0$.

\section{results}
Before numerical calculations, we consider the parameters of IQMDD
model at first. To guarantee our model can be used not only to
explain the saturation properties of nuclear matter, but also to
describe the quark deconfinement phase transition, we fix our
parameters as those in Ref. [6]. Hereafter we fix the parameters as:
 at zero temperature, the bag constant $B$=174 MeV$fm^{-3}$, the masses of
 $\omega$- meson and $\sigma$- meson $m_\omega = 783$ MeV and
 $m_\sigma=509$ MeV, the coupling constant $f = 5.45$ and $g=3.37$
respectively and the mass of nucleon $M_N = 939$ MeV,
 $b=-8400$ MeV. As was shown in Ref. [6], this set of parameters can give us
 reasonable properties of nuclear matter, including the value of
 saturation point, the equation of state and the compressibility.
  we  will prove  in this  section that we  can also obtain
  a reasonable quark deconfinement  critical temperature $T_C$ by means
  of this set of parameters.

The set of coupled differential equations (19), (20) with boundary
conditions Eq. (22) can be solved numerically at zero temperature.
Our results at zero temperature are shown in Figs. (1-4). The
variation of the scalar field $\sigma$ as a function of the radius r
is presented in Fig. 1. We see that the value of $\sigma$ inside the
hadron is very different from that of outside: inside $\sigma$ is
less than zero, and outside $\sigma \rightarrow \sigma_v$. The
transitional values of $\sigma$ field through the surface from
inside to outside is abrupt. The curve in Fig. 1 is very similar as
that given by Refs. [17, 22] where the $\omega$- meson is omitted.
This is reasonable because  the main  contribution to form a soliton
bag comes from the nonlinear $\sigma$ field. The variation of wave
function of quark  field is shown in Fig. 2 and Fig. 3. In Fig. 2,
we plot the curves of quark wave function $u, v$ vs. $r$
respectively. In Fig. 3, we plot the curve of quark density
$u^2-v^2$ vs. $r$. The soliton bag is exhibited in this figure
transparently. The curve of $\omega$- field vs. $r$ given by Eq.
(21) is shown in Fig. 4. This solution is obtained from Green
function method and Eq. (23). We see from this Figure that the curve
of $\omega$- field
 decays considerably when $r$ is large than the soliton bag
radius.

Now we turn to investigate the case of finite temperature. The
effective potential at finite temperature can be obtained by
numerical calculations from the set of Eqs. (24-27). The curves of
effective potential $U(\sigma; T; \mu; V_\omega)$ at different
temperatures $T=0$ MeV, $T=80$ MeV, $T=127$ MeV and $T=204$ MeV are
shown in Fig. 5 respectively. The shape of the effective potential
confirms that a first order phase transition will take place [22,
25-27]. We see from Fig. 5 there are two vacua where one corresponds
to the physical vacuum and the other corresponds to the false vacuum
when $T < 127$ MeV. But when $T$ increases to a critical temperature
$T_C=127$ MeV, these two vacua degenerate and the values of bag
constant tends to zero.

To show this behaviour more clearly, we plot the bag constant $B$
vs. $T$ curve by solid curve in Fig. 6, we see that the bag constant
$B$ decreases as temperature $T$ increases. When $T$ approaches to
$T_C$, $B$ approaches to zero and quark deconfined phase transition
happens.

When temperature increases to the regions 127 MeV $< T <204$ MeV, as
shown in Fig. 5 the physical vacuum becomes instable and the
purturbative vacuum becomes stable. The soliton solution tends to
disappear [22]. When $T$ approached to 204 MeV, the effective
potential becomes to have an unique minimum only. Such potential no
longer ensures the existence of the soliton bag anymore because the
spontaneous breaking symmetry has been restored. We have zero
solution $\sigma=0$ in the regions $T \geq 204$ MeV only.

To illustrate our soliton solutions more transparently, we plot the
soliton solutions by fixed the temperature $T=0$ MeV, $T=80$ MeV and
$T=125$ MeV in Fig. 7. It is shown in Fig. 7 that the curves stretch
slowly to infinite with increasing temperature. The radius of
soliton increases and the skin of  the soliton becomes thicker when
the  temperature increases. It is of course very reasonable.

All characters of the soliton solutions found by IQMDD model are
almost the same as those given by Ref. [22] where the $\omega$ field
has not been take into account. But we hope to emphasize although
the $\omega$ field will not affect the main characters of the
soliton solutions, but it will change their detailed behavior. To
address the effect of $\omega$ field on the deconfinement
transition, as an example, we plot the $B(T)$ curve without $\omega$
field by dotted line in Fig. 6. We find the critical temperature
from the condition $B(T_C)=0$ changes to $T_C=140$ MeV which is
higher than 127 MeV where the $\omega$ field existed. If we notice
that the interaction of $qq\omega$ is repulsive, and then it can
help to deconfine quarks,  the decrease of $T_C$ due to $\omega$-
field is natural.

 \section{Summary and discussion}
By using the finite temperature quantum field theory, we calculate
the finite temperature effective potential  and extend our previous
discussions to finite temperature. It is shown that the IQMDD model
can not only  describe the saturation properties of nuclear matter,
but also  explain the quark deconfinement phase transition
successfully. We have shown the soliton solution curves for
different temperatures and found the critical temperature $T_C=127$
MeV. The $\omega$- field in IQMDD model is important. The repulsive
$qq\omega$ interaction  plays the central role to describe the
saturation properties, and affect the critical temperature
remarkably. Comparing to naive QMC model, the advantage of IQMDD
model is obvious because instead of MIT bag, a Friedberg-Lee soliton
bag exists in this model.

Of course, there are still shortcomings in the IQMDD model. For
example, the interactions between quarks and mesons are still
isospin independent. The chiral phase transition cannot be discussed
by this model because it is lack of chiral symmetry. To Overcome
these shortcomings, we hope to add the $\rho$- meson and $\pi$-
meson in IQMDD model in the near future.

\begin{acknowledgments}
This work is supported in part by the National Natural Science
Foundation of People's Republic of China. C Wu  is extremely
grateful to Dr. H Mao for useful discussions and correspondence.
\end{acknowledgments}

\begin{figure}[tbp]
\includegraphics[width=14cm,height=20cm]{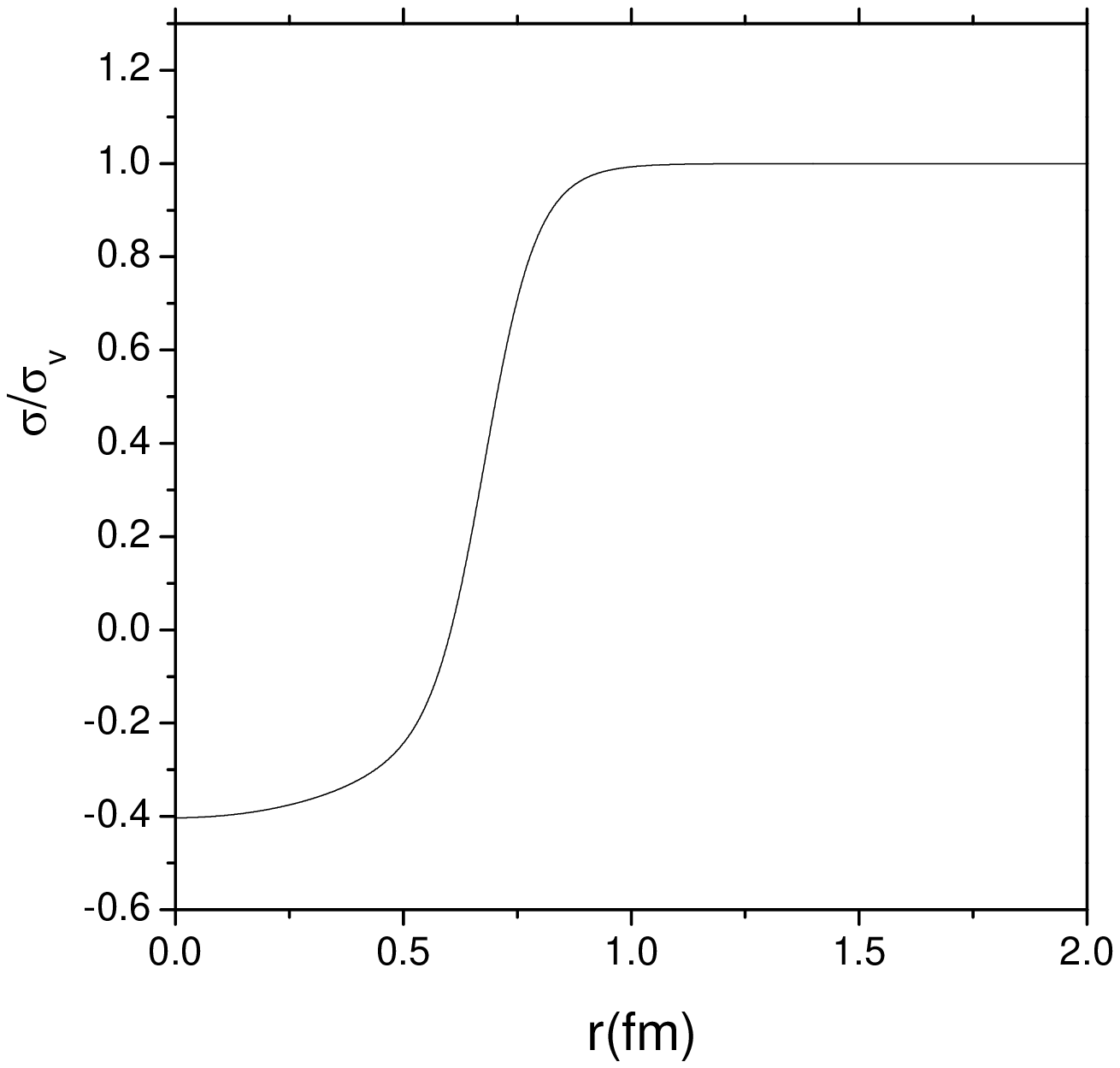}
\caption{The $\sigma$ field  as a functions of r at zero
temperature.  }
\end{figure}

\begin{figure}[tbp]
\includegraphics[width=14cm,height=20cm]{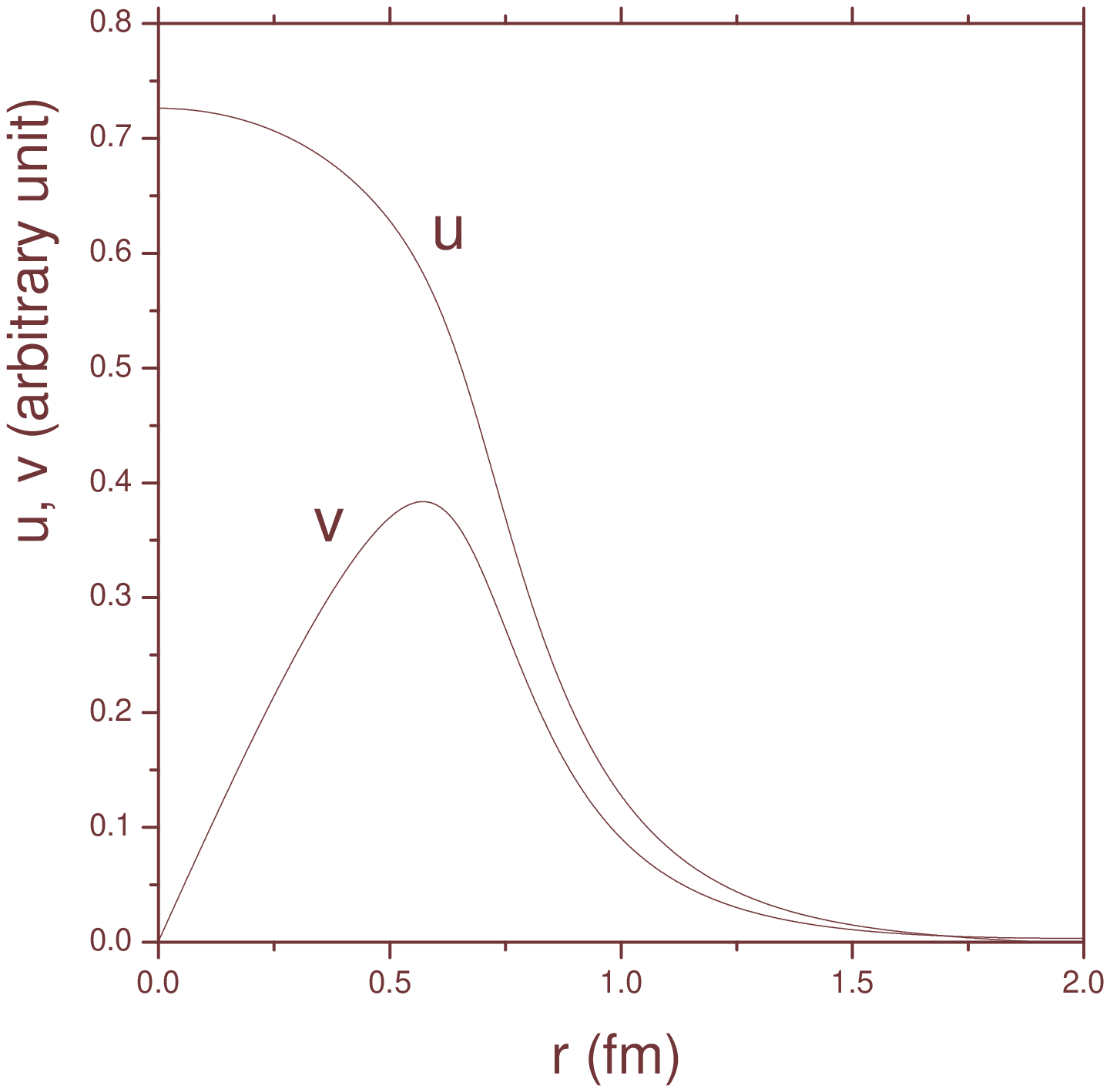} \caption{Quark
wave functions in arbitrary unit as a functions of r.}
\end{figure}

\begin{figure}[tbp]
\includegraphics[width=14cm,height=20cm]{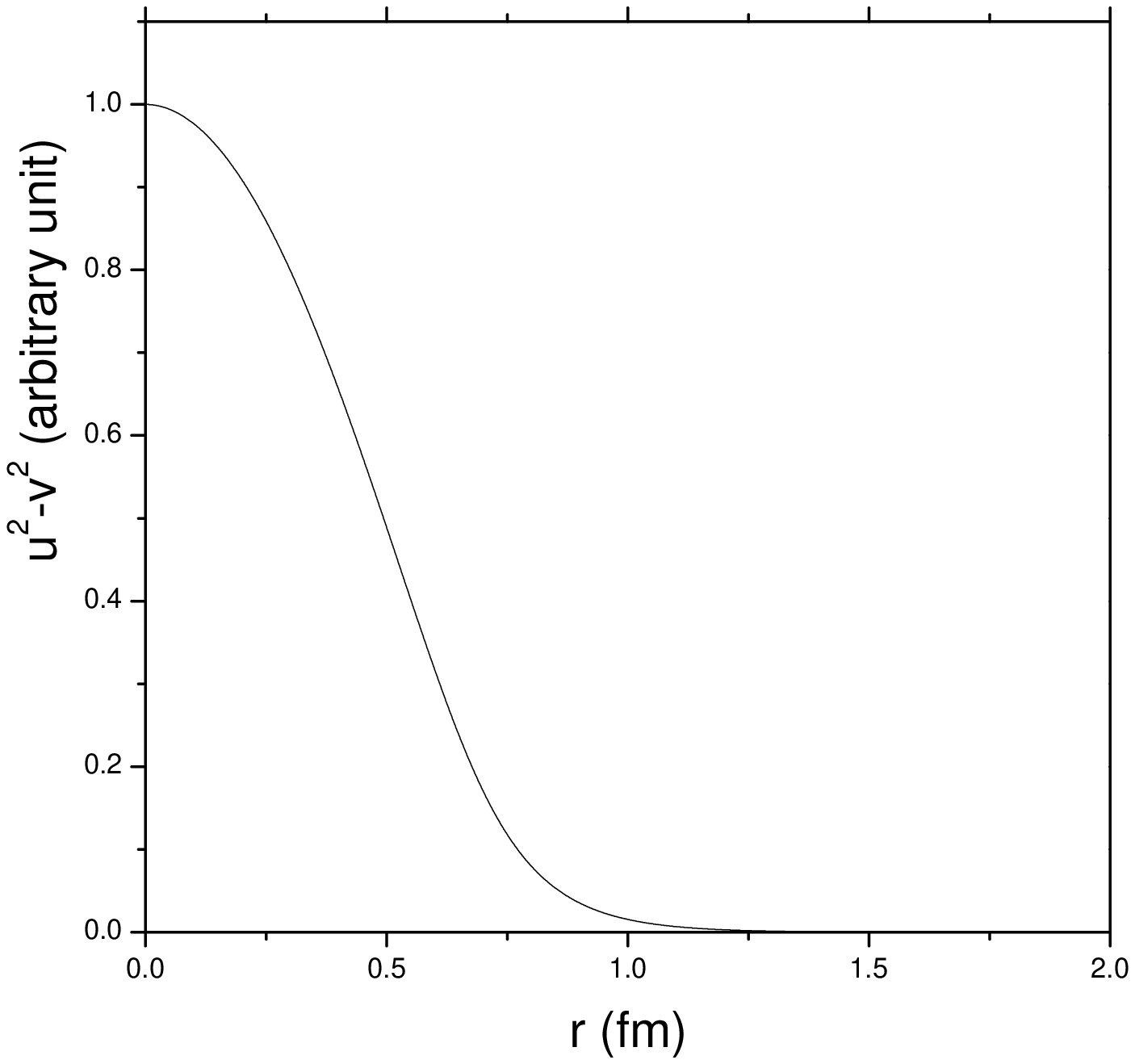}
\caption{The quark density $u^2(r)-v^2(r)$ in arbitrary unit as a
functions of r.}

\end{figure}

\begin{figure}[tbp]
\includegraphics[width=14cm,height=20cm]{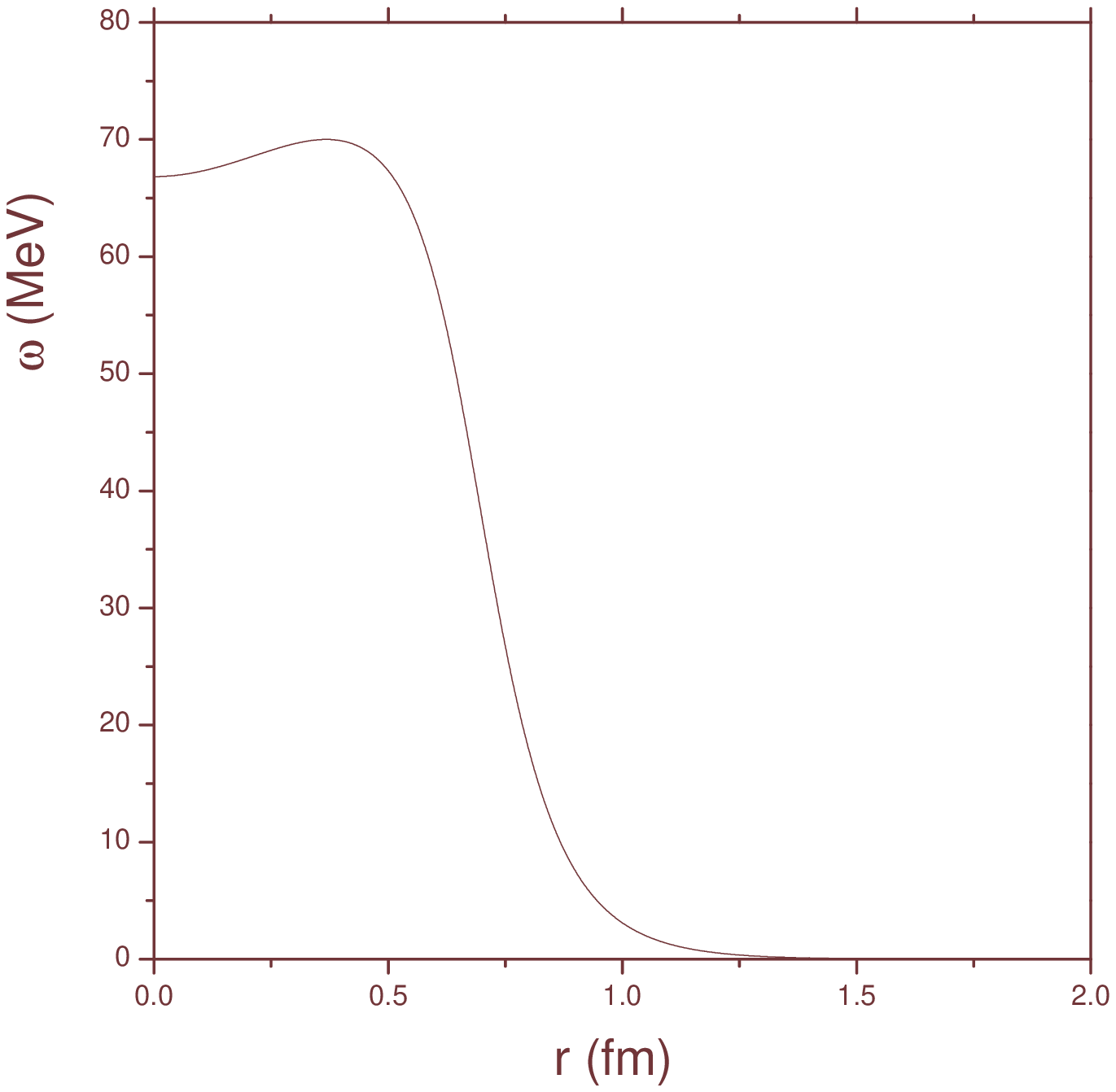}
\caption{The $\omega$ field  as a functions of r.}

\end{figure}
\begin{figure}[tbp]
\includegraphics[width=14cm,height=20cm]{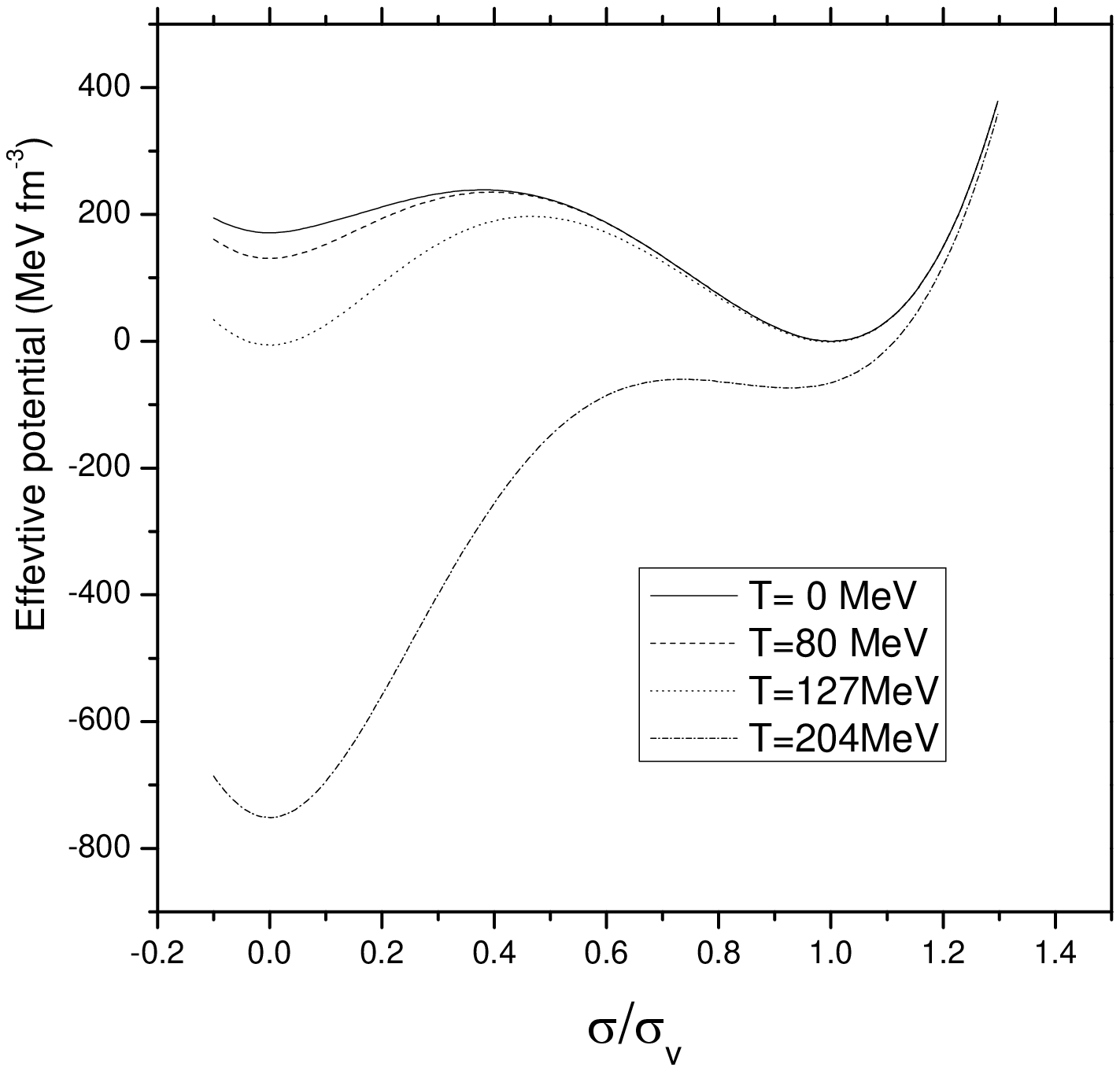}
\caption{The temperature-dependent  effective potential.}
\end{figure}

\begin{figure}[tbp]
\includegraphics[width=14cm,height=20cm]{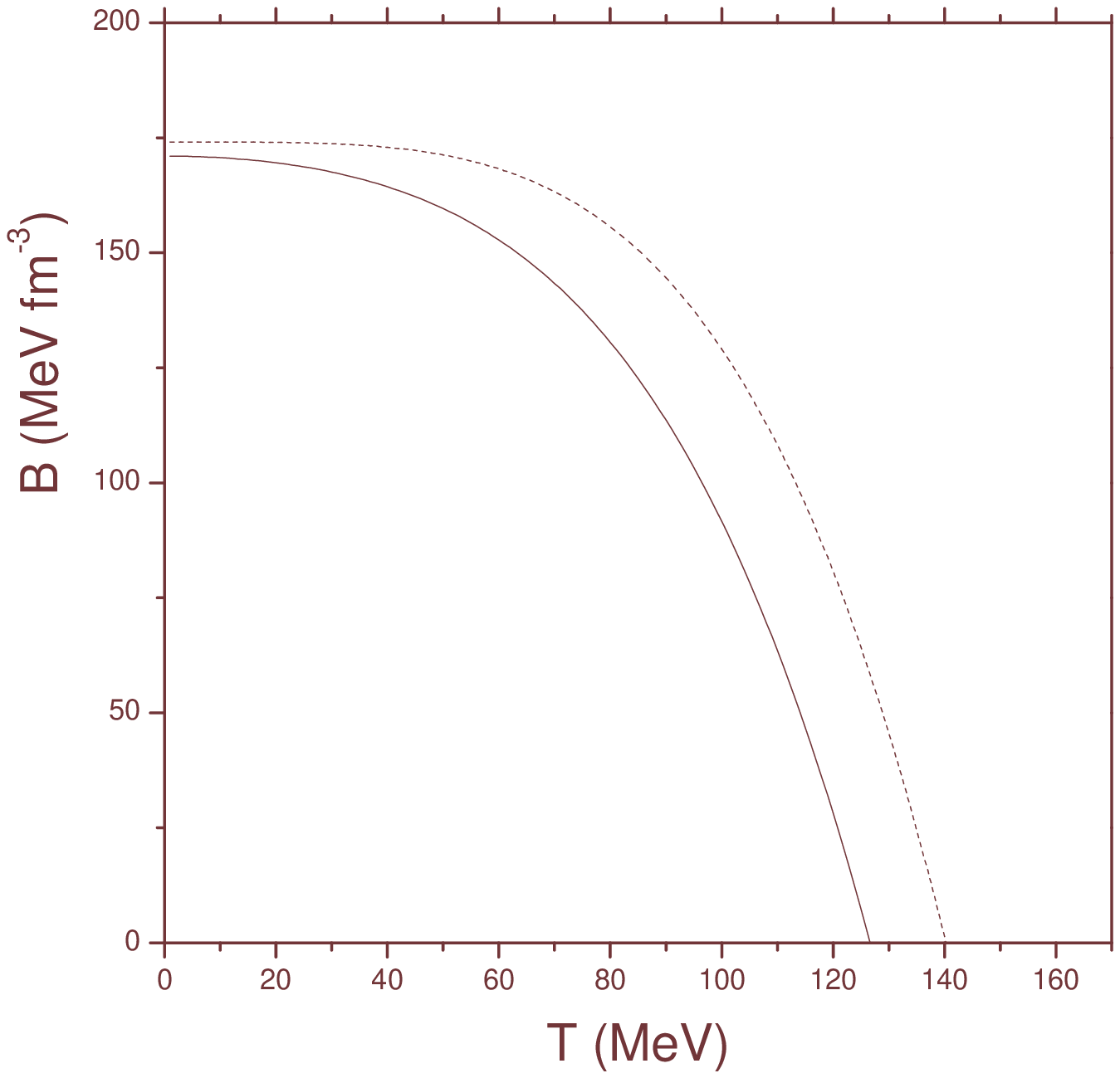}
\caption{The bag constant B(T) as functions of T, the solid line
corresponds to IQMDD model with $\omega$-meson and the dotted line
corresponds to the same model without $\omega$-meson.}
\end{figure}

\begin{figure}[tbp]
\includegraphics[width=14cm,height=20cm]{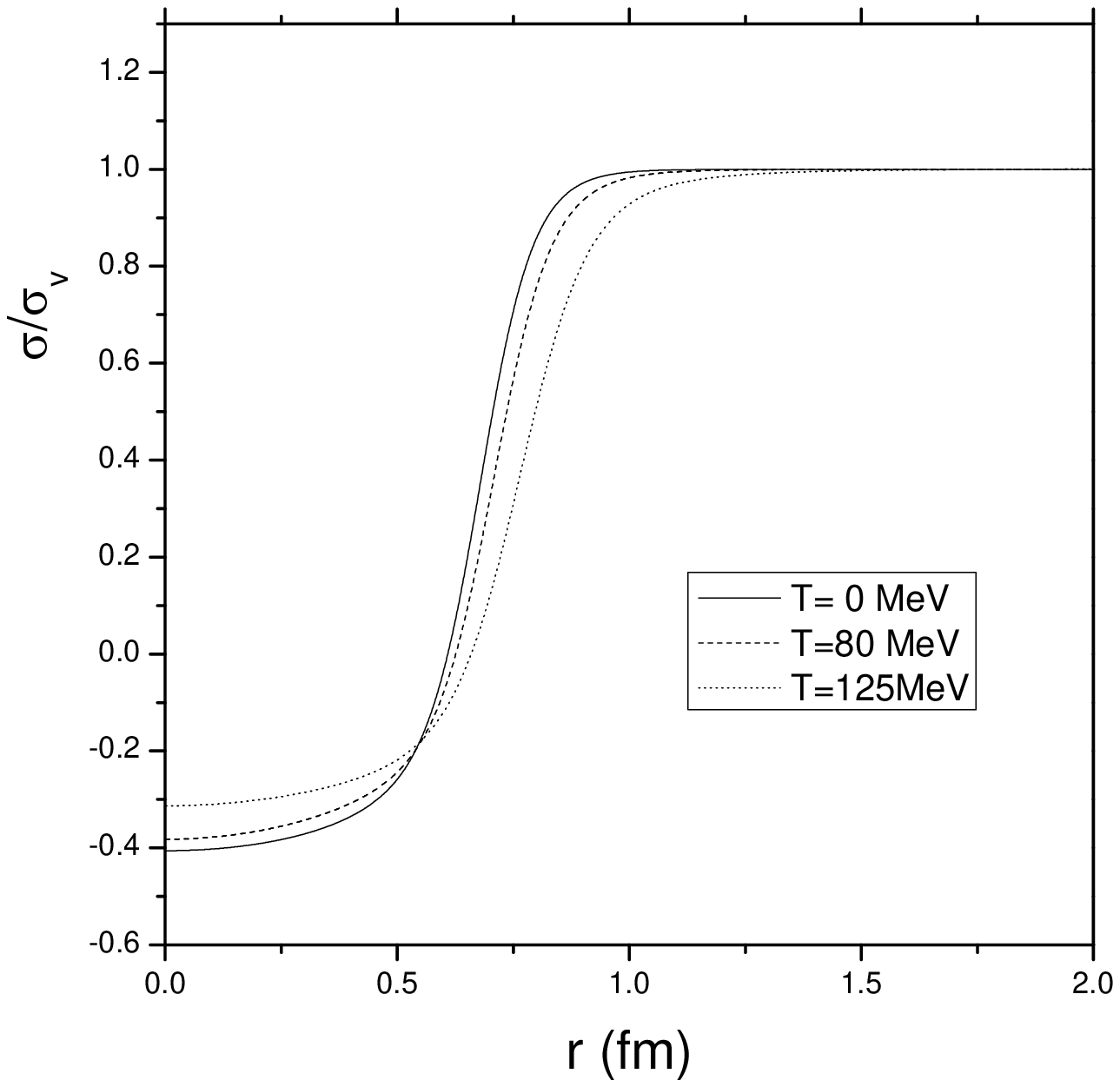}
\caption{The soliton solutions for different temperature T=0 MeV, 80
MeV, 125 MeV}
\end{figure}

\end{document}